\documentclass[a4paper,twoside]{article}
\usepackage[T1]{fontenc}
\usepackage[latin1]{inputenc}
\usepackage[frenchb,english]{babel}
\usepackage{graphicx}
\usepackage[noindentafter,pagestyles]{titlesec}
\usepackage{textcomp}
\usepackage{dcolumn}
\usepackage{amsmath}
\usepackage{txfonts}
\usepackage{fancyhdr}
\usepackage{threeparttable}
\usepackage{xspace}
\usepackage[superscript]{cite}
\titleformat{\section}{\large\bfseries}{\thesection.}{.3em}{}
\titlespacing*{\section}{\leftmargini}{*3}{*3}
\titleformat{\subsection}{\bfseries}{\thesubsection}{.3em}{}
\titlespacing*{\subsection}{0pt}{*3}{*3}
\makeatletter
\def\@maketitle{%
  \newpage
  \null
  \vskip 2em%
  \begin{center}%
  \let \footnote \thanks
    {\fontsize{18}{22}\fontseries{b}\selectfont \@title \par}%
    \vskip 1.5em%
    {\normalsize
      \lineskip .5em%
      \begin{tabular}[t]{c}%
\@author
      \end{tabular}\par}%
    \vskip 1em%
    {\large \@date}%
  \end{center}%
  \par
  \vskip 1.5em}
\renewenvironment{abstract}{%
\if@twocolumn
\section*{\abstractname}%
\else \quotation \noindent{\bfseries\large
\abstractname\vspace*{.3ex}\par} \fi}
{\if@twocolumn\else\endquotation\fi} \makeatother

\fancypagestyle{plain}{%
\fancyhf{}
\fancyhead[R]{\small \MakeUppercase{ }}%
\fancyfoot[L]{\small C.~J.~Greenshields and
J.~M.~Reese, 1st June 2007}%
}   \renewcommand{\Pr}{\mbox{\textit{Pr}}}
\newcommand{\Ma}{\mbox{\textit{Ma}}}
\newcommand{\Kn}{\mbox{\textit{Kn}}}
\newcommand{\Sc}{\mbox{\textit{Sc}}}
\newcommand{\im} {\mathrm{i}}

\newcommand{\stress} {\vec{P}}
\newcommand{\vStress} {\vec{T}}
\newcommand{\D} {\vec{D}}
\newcommand{\I} {\vec{I}}
\newcommand{\U} {\vec{u}}
\newcommand{\Um} {\U_{m}}
\newcommand{\J} {\vec{j}}

\newcommand{\etc} {etc.\@\xspace}
\newcommand{\ie} {i.e.\@\xspace}
\newcommand{\eg} {e.g.\@\xspace}

\providecommand\bnabla{\boldsymbol{\nabla}}
\providecommand\bcdot{\boldsymbol{\cdot}}
\renewcommand{\vec}[1] {\boldsymbol{#1}}
\newcommand{\dprod}{\bcdot}
\renewcommand{\div}{\bnabla\dprod}
\newcommand{\grad}{\bnabla}

\newcommand{\DDt}[1]{\frac{D #1}{D t}}
\newcommand{\ddt}[1]{\frac{\partial #1}{\partial t }}
\newcommand{\tr} {\mbox{tr}}
\newcommand{\trans}[1] {\ensuremath{#1^{\mathrm{T}}}}
\newcommand{\devs}{\textrm{dev}}
\newcommand{\symm}{\textrm{symm}}
\newlength{\skewslength}
\newlength{\skewsheight}

\providecommand{\url}[1]{\texttt{#1}}
\begin{document}
\title{The structure of hypersonic shock waves using Navier-Stokes equations
modified to include mass diffusion}
\author{\itshape C.~J.~Greenshields and J.~M.~Reese\\
\itshape Department of Mechanical Engineering, University of
Strathclyde\\
\itshape Glasgow G1 1XJ, United Kingdom}

\date{}
\maketitle
\begin{abstract}
\noindent %
Howard Brenner\cite{Brenner:2004,Brenner:2005a,Brenner:2005b} has
recently proposed modifications to the Navier-Stokes equations that
relate to a diffusion of fluid volume that would be significant for
flows with high density gradients.  In a previous
paper\cite{Greenshields&Reese:2007}, we found these modifications
gave good predictions of the viscous structure of shock waves in
argon in the range Mach 1.0--12.0 (while conventional Navier-Stokes
equations are known to fail above about Mach 2).  However, some
areas of concern with this model were a somewhat arbitrary choice of
modelling coefficient, and potentially unphysical and unstable
solutions.  In this paper, we therefore present slightly different
modifications to include molecule mass diffusion fully in the
Navier-Stokes equations.  These modifications are shown to be stable
and produce physical solutions to the shock problem of a quality
broadly similar to those from the family of extended hydrodynamic
models that includes the Burnett equations.  The modifications
primarily add a diffusion term to the mass conservation equation, so
are at least as simple to solve as the Navier-Stokes equations;
there are none of the numerical implementation problems of
conventional extended hydrodynamics models, particularly in respect
of boundary conditions.  We recommend further investigation and
testing on a number of different benchmark non-equilibrium flow
cases.
\end{abstract}

\section{Background}
\label{sec:background}
A parameter which indicates the extent to which a local region of
flowing gas is in thermodynamic equilibrium is the Knudsen number:
\begin{equation}
    \Kn
  = \frac{\lambda}{L}
  \approx
    \frac{\lambda}{\rho} \left|\grad\rho\right|,
    \label{eq:Knudsen}
\end{equation}
where $\lambda$ is the mean free path of the gas molecules, $L$ is a
characteristic length of the flow system and $\rho$ is a
characteristic mass density. As $\Kn$ increases, \eg for vehicles
travelling at hypersonic speeds or at high altitudes, the departure
of the gas from local thermodynamic equilibrium increases, and the
notion of the gas as a continuum-equilibrium fluid becomes less
valid.  The Navier-Stokes equations (with standard no-slip boundary
conditions) are, for example, typically confined to cases where $\Kn
\lesssim 0.01$.  Their underlying constitutive laws for viscous
stress tensor $\vStress$ and heat flux $\J_{e}$, \ie Newton's law
and Fourier's law respectively, may be derived from the Boltzmann
equation using the classical Chapman-Enskog expansion in $\Kn$ to
first order. {\em Extended}, or {\em modified}, {\em hydrodynamics}
models, such as the Burnett equations, are based on Chapman-Enskog
expansions to higher orders in an attempt to extend the range of
applicability of the continuum-equilibrium fluid model into the
so-called `intermediate-$\Kn$' (or `transition-continuum') regime
where $0.01\lesssim\Kn\lesssim 1$.  Extended expressions for
$\vStress$ and $\J_{e}$ include the same terms to first order in
$\Kn$ contained in Newton's and Fourier's laws respectively, but
with the addition of numerous, more complex terms that make them
notoriously unstable and costly to solve.

In 2004\cite{Brenner:2004}, Howard Brenner proposed that the
velocity $\U$ appearing in Newton's viscosity law is generally
different from the mass velocity $\Um$ appearing in the mass
conservation equation:
\begin{equation}
    \ddt{\rho}
  + \div ( \rho\Um )
  = 0.
    \label{eq:massContinuity}
\end{equation}
He subsequently related the two velocities by a \emph{diffusive
volume flux density} $\J_{v} = \U - \Um$ and proposed a constitutive
model relating $\J_{v}$ to $\grad\rho$, similar to Fick's law of
mass diffusion\cite{Brenner:2005a}. From this he derived
modifications to the Navier-Stokes equations in which the governing
transport equations of mass, momentum and energy remained unchanged
but the constitutive equations for $\vStress$ and $\J_{e}$ are
augmented by additional terms\cite{Brenner:2005b}.  The resulting
equations have the form of an extended hydrodynamics model, albeit
one much less complex than the family of hydrodynamics models that
includes Burnett, Grad, \etc

The viscous structure of shock waves in gases provides an obvious
test case for Brenner's modifications since they become increasingly
significant as $\grad\rho$ increases.  While it is accepted that the
Navier-Stokes equations fail in nearly every respect in predicting
correct shock structures above about Mach 2, they do reproduce the
trends in experimental and molecular dynamics simulation data
significantly better when Brenner's modifications are included,
delivering an excellent match in the case of the inverse density
thickness~\cite{Greenshields&Reese:2007}. It is of some concern,
however, that shock solutions are nonphysical when the
proportionality coefficient $D_{v}$, used in the constitutive model
for volume diffusion, exceeds approximately the kinematic viscosity %
$\nu = \mu/\rho$, where $\mu$ is the dynamic viscosity.  Furthermore
solutions were shown to be unstable when $D_{v} \gtrsim 1.45\nu$.
The imposed limit on $D_{v}$ effectively dictates the choice of
$D_{v} = \nu$ for which there is apparently no strong physical
justification.

Nevertheless, the results do partially support Brenner's original
hypothesis --- which is undoubtedly viewed with scepticism by some
because it challenges the fundamental equations of fluid mechanics.
Particular areas of criticism are that the underlying theory lacks a
sound physical basis and is based on the notion of a diffusive
volume flux which is somewhat difficult to conceptualise and for
which constitutive models and their coefficients are untested.
However, support can be found in the phenomenological GENERIC theory
presented by Hans Christian \"{O}ttinger\cite{Oettinger:2005}.
\"{O}ttinger questions why a diffusive transport term exists for
both energy and momentum but not mass in the conventional
Navier-Stokes equations and argues ``something is missing'', namely
the ability of mass diffusion to produce entropy.  When dissipative
terms associated with mass density are identically zero, the GENERIC
formulation arrives at the standard Navier-Stokes equations but, by
including non-zero terms, a revised set of governing equations is
derived that includes two velocities, similar to $\U_{m}$ and $\U$.
The modifications are simply due to mass diffusion rather than the
difficult concept of a diffusive volume flux.

Brenner subsequently adopted\cite{Brenner:2006} the equations of
\"{O}ttinger which differ from his original modifications
particularly in that $\U$ appears not only in Newton's viscosity law
but in the definition of momentum density itself. The purpose of
this present paper is to provide additional argument in favour of
the inclusion of mass diffusion and to examine its impact on the
governing equations in detail.  We assess the stability of the
underlying equations and their ability to predict the structure of
shock waves.

\section{Mass diffusion and conservation}
\label{sec:massDiffusivity}
Thermal agitation causes molecules to travel from one region of a
gas to another.  Inequalities in molecular distribution and thermal
velocity tend to be smoothed by an inevitable net migration of
molecules towards regions of lower molecular concentration and/or
temperature.  This is the process by which mass diffuses and, in the
simplest case of a single specie gas, a diffusive flux $\J_{d}$
occurs in the direction of negative density gradient, expressed
through Fick's law simply as $\J_{d} = -D_{m} \grad \rho$ where
$D_{m}$ is the coefficient of mass diffusion (self-diffusion, in the
case of a single specie gas).  The conventional governing equations
clearly omit the process of mass diffusion due to \emph{net}
migration of molecules by thermal agitation because they do not
contain Fick's or any other constitutive model in the equation of
conservation of mass.  This is important in extended hydrodynamic
modelling of hypersonic flows because the omission becomes more
significant as $\grad\rho$ and, thus, $\Kn$ become larger.

Modelling of mass diffusion is, of course, a common feature in the
analysis of multicomponent fluid systems.  The usual approach is to
retain the conservation equation for total fluid mass given by
(\ref{eq:massContinuity}) and create additional conservation
equations for the mass of individual gas species that each include a
diffusive mass flux term\cite{Bird&al:2002}. However, for $N$ gas
species, only $N - 1$ equations of specie mass conservation are
independent because the sum of all $N$ equations gives
(\ref{eq:massContinuity}).  This means that mass diffusion is not
modelled for one of the species, a statement that applies to the
case of a single specie gas described in the previous paragraph. It
happens because (\ref{eq:massContinuity}) \emph{defines} $\rho\Um$
to be the \emph{total} mass flux density, \ie the sum of the bulk,
advective mass flux of the fluid and the net sum of diffusive mass
flux of all consitutents of the fluid.  In other words, $\Um$
constitutes a mean, or local mass average, velocity of all
consitutents of the fluid\cite{Bird&al:2002} of which the advective
velocity is only a part.

The inability to model net mass diffusion and associated
irreversible energy dissipation are not the only worrying
consequences of combining advective and net diffusive fluxes into a
single velocity $\Um$.  It has been argued that, by doing this,
Fick's law for a constituent is applied relative to the net flux of
all constituents when it is only applicable relative to a frame of
reference external to the fluid\cite{Corey&Auvermann:2003}.  Also,
velocity associated with mass diffusion has questionable physical
significance because where the concentration of a given
specie~$\rightarrow 0$, its diffusion velocity~$\rightarrow
\infty$\cite{Mills:1998}.

Instead, let us split the total mass flux density $\rho\Um = \rho\U
+ \J_{d}$ where $\U$ is termed the advective velocity. If the
diffusive flux is modelled by Fick's law, this leads to the
following mass continuity equation:
\begin{equation}
    \ddt{\rho}
  + \div ( \rho\U )
  - \div ( D_{m} \grad\rho )
  = 0.
    \label{eq:modifiedMassContinuity}
\end{equation}
This is the form of mass conservation equation proposed by
\"{O}ttinger expressed in terms of $\U$, not $\Um$.  It is
interesting to observe that the ratio $R_{dc}$ of diffusive mass
flux to advective mass flux can be expressed as
\begin{equation}
    R_{dc}
  = \left|\frac{D_{m}\grad\rho}{\rho\U}\right|
  = \frac{1}{A_{\lambda}\sqrt{\gamma}\Sc}\frac{\Kn}{\Ma},
    \label{eq:ratioMassFluxes}
\end{equation}
where: the Mach number of the flow $\Ma = |\U|/c$ with $c$ the speed
of sound; the Schmidt number $\Sc = \nu/D_{m}$; $\gamma$ is the
ratio of specific heats at constant pressure and volume; $\Kn$ is
based on a Maxwellian mean free path $\lambda_{M} =
A_{\lambda}\sqrt{\gamma}\nu/c$, with $A_{\lambda} =
16/(5\sqrt{2\pi}) \approx 1.28$. For argon gas, $\gamma = 5/3$ and
the coefficient of self-diffusion of mass $D_{m} \approx
1.32\nu$,\cite{Winn:1950,Chapman&Cowling:1970,Bird:1970} so $\Sc =
0.76$ and then $R_{dc} \approx 0.8\Kn/\Ma$. In a planar shock in
argon at Mach 4 upstream, $\Kn \approx 0.275$ (see
figure~\ref{fig:MachVsInvShockThickness}) and $\Ma \approx 1.0$
local to the midpoint across the density profile, so $R_{dc} \approx
22\%$. The omission of mass diffusion from the governing equations
will clearly lead to error at such a high $R_{dc}$.  It could also
be expected that $R_{dc}$ is high in regions of low speed and
moderate density gradient such as boundary layers and wakes, both
regions where the departure from non-equilibrium behaviour is most
pronounced in hypersonic flows\cite{Lofthouse&al:2007}.

\section{Momentum and energy conservation}
\label{sec:momentumEnergyConservation}
Brenner's original hypothesis was that Newton's viscosity law should
be expressed in terms of $\U$ %
, not $\Um$. %
Along with his own supporting analytical and experimental evidence,
there is also the argument that velocity in Newton's viscosity law
represents deformation of fluid volume and therefore cannot be based
on a \emph{mass average} velocity $\Um$.\cite{Corey&Auvermann:2003}.
The argument that velocity relating to mass diffusion has no
physical significance\cite{Mills:1998} effectively precludes the use
of $\Um$ in Newton's viscosity law.

\"{O}ttinger additionally defines momentum density as $\rho\U$, not
$\rho\Um$, so that momentum relates purely to advective mass flux,
not diffusive mass flux.  This seems reasonable given that the
advective flux is caused by mean translatory motion of molecules,
associated with mechanical energy whereas the diffusive mass flux is
caused by random motion of molecules associated with thermal energy.
At the very least, if some momentum is attributed to mass diffusion,
it must relate to a separate driving force independent of viscous
forces associated with advective
momentum.\cite{Corey&Auvermann:2003}  The resulting equation can be
alternatively viewed as the governing equation for momentum in the
absence of a (net) diffusive mass flux.  Following these arguments,
the momentum equation is as in the standard governing equations
(ignoring body forces) but expressed in terms of $\U$, not $\Um$:
\begin{equation}
    \rho\DDt{\U}
  = \ddt{(\rho\U)}
  + \div ( \rho\U_{m}\U )
  = \div \stress,
    \label{eq:modifiedMomentumContinuity}
\end{equation}
where the stress tensor $\stress =  \vStress - p\I $ (defined as
positive in tension), $p$ is pressure and $\I$ the unit tensor.
Newton's law is expressed as
\begin{math}
\vStress = 2\mu\,\devs (\D) + \kappa\,\tr(\D)\I
\label{eq:NewtonianFluid}
\end{math}
where $\kappa$ is the bulk viscosity, the deformation gradient
tensor
\begin{math}
   \D
 \equiv
   \symm(\grad \U)
 \equiv (1/2) \left[ \grad\U + \trans{(\grad \U)} \right]
\end{math}
and its
deviatoric component $\devs (\D) \equiv \D - (1/3)\,\tr(\D)\I$.
Note that the material derivative $D/Dt$ is decomposed into the
local rate of change $\partial/\partial t$ and the convective rate
of change based on the local velocity of a fluid element that both
advects and diffuses, \ie $\Um$.

The derivation of a conservation equation for energy within a
continuum framework that includes mass diffusion is more
challenging.  The energy equation derived using
GENERIC\cite{Oettinger:2005}, for example, contains an unconstrained
phenomenological parameter that has to be determined by theory or
simulation and confirmed by experiment.  Derivation through physical
argument alone requires careful accounting of contributions of mass
diffusion to mechanical and thermal energies and associated work. As
a first approximation, we equate the rate of change of total energy
to mechanical and thermodynamic energy fluxes (ignoring internal
sources):
\begin{equation}
    \rho\DDt{E}
  =
    \ddt{(\rho E)}
  + \div(\rho\Um E)
  =
    \div(\stress\dprod\U)
  - \div\J_{e},
    \label{eq:totalEnergy}
\end{equation}
where the mechanical energy flux density ($\stress\dprod\U$) relates
to the advective velocity $\U$ only and heat energy flux is
attributed to conduction only by Fourier's law $\J_{e} = - k\grad T$
where $k$ is the thermal conductivity. The total energy per unit
mass $E$ represents all mechanical and thermal energy contributions.
Initial simulations of the shock structure problem showed the
temperature-density separation, discussed in
section~\ref{sec:temperatureDensitySeparation}, was strongly
underpredicted when the $E$ included a kinetic energy due to
advective mass flux only. Instead, much better predictions were
obtained when the kinetic energy was due to the total mass flux such
that
\begin{math}
   E
 = e
 + |\Um|^{2}/2,
\end{math}
suggesting that net mass diffusion contributes to an additional
source of energy beyond the internal energy.

\section{Stability analysis}
\label{sec:stabilityAnalysis}
The same stability analysis was undertaken on the set of governing
equations proposed in this work that previously highlighted
limitations of Brenner's original
modifications\cite{Greenshields&Reese:2007}.  Following the
procedures described previously
\cite{Zhong&al:1991,Struchtrup&Torrilhon:2003,Greenshields&Reese:2007},
it is first assumed that the gas is monatomic and calorically
perfect with $\gamma = 5/3$, Prandtl number $\Pr = [\gamma R/(\gamma
- 1)](\mu/k) = 2/3$, where $R$ is the gas constant, and $\kappa =
0$. The governing equations from sections~\ref{sec:massDiffusivity}
and \ref{sec:momentumEnergyConservation} are linearised in
1-dimension to produce the following non-dimensionalised
perturbation equations:
\begin{equation}
    \frac{\partial \phi}{\partial t^{\prime}}
  +
    \begin{bmatrix}
        0  & 1 &  0 \\
        1 & 0 & 1 \\
        0 & \frac{2}{3} & 0
    \end{bmatrix}
    \frac{\partial \phi}{\partial x^{\prime}}
  + \frac{\partial}{\partial x^{\prime}}
    \left\{
        \begin{array}{c}
            c^{\prime} \\ \sigma^{\prime} \\ q^{\prime}
        \end{array}
    \right\}
  =
    0 ,
    \label{eq:pertubation}
\end{equation}
where
\begin{equation}
    c^{\prime}
  =
  - \frac{1}{\Sc}
    \frac{\partial \rho^{\prime}}
         {\partial x^{\prime}},
\quad
    \sigma^{\prime}
  =
  - \frac{4}{3} \frac{\partial u^{\prime}}{\partial x^{\prime}}
\quad
\textrm{and}
\quad
    q^{\prime}
  =
  - \frac{5}{2} \frac{\partial T^{\prime}}{\partial x^{\prime}}.
\end{equation}
We assume a solution to (\ref{eq:pertubation}) of the form
\begin{equation}
    \phi
  =
    \tilde{\phi}
    \exp\left\{\im(\omega t^{\prime} - k x^{\prime})\right\} ,
    \label{eq:perturbationSolution}
\end{equation}
where $\tilde{\phi}$ is the amplitude of the wave, $\omega$ is its
frequency and $k$ its propagation constant.
Equations~(\ref{eq:pertubation}) to (\ref{eq:perturbationSolution})
can be combined to produce a set of linear algebraic equations of
the form
\begin{equation}
    \mathcal{A}(\omega,k) \tilde{\phi} = 0 ,
    \label{eq:linearAlgebraicPerurbation}
\end{equation}
for which non-trivial solutions require
\begin{equation}
    \det[\mathcal{A}(\omega,k)] = 0 .
    \label{eq:linearAlgebraicPerurbationDeterminant}
\end{equation}
For our modified governing equations,
(\ref{eq:linearAlgebraicPerurbationDeterminant}) yields the
following characteristic equation:
\begin{equation}
    6\im  \omega^{3}
  + (23 + 6\Sc^{-1})k^{2} \omega^{2}
  - [10 k^{2} + (20 + 23 \Sc^{-1})k^{4}]\im \omega
  - [(15 + 4\Sc^{-1}) k^{4} + 20\Sc^{-1}k^{6}]
  = 0 .
    \label{eq:BNScharacteristics}
\end{equation}
If a disturbance in space is considered as an initial-value problem,
$k$ is real and $\omega = \omega_{r} + \im\omega_{i}$ is complex.
The form of (\ref{eq:perturbationSolution}) indicates that stability
then requires $\omega_{i} \ge 0$.  If a disturbance in time is
considered as a problem of signalling from the boundary, $\omega$ is
real and $k = k_{r} + \im k_{i}$ is complex. For a wave travelling
in the positive $x$ direction, $k_{r} > 0$, and stability then
requires that $k_{i} < 0$.  For a wave travelling in the negative
$x$ direction, the converse is true: $k_{r} < 0$ and stability
requires $k_{i} > 0$.

\begin{figure}[t]
    \centering
    \input{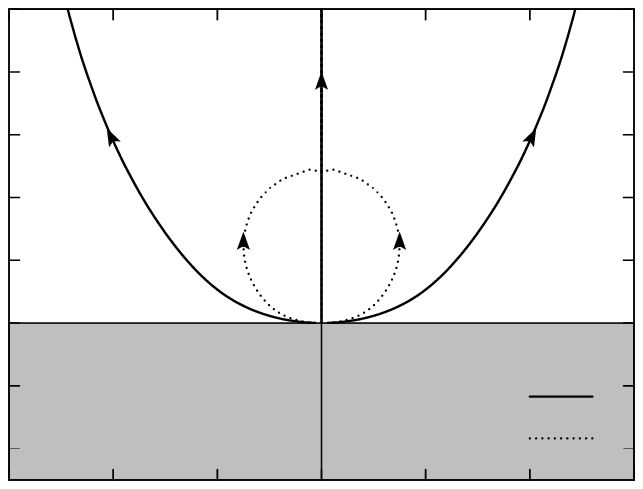}
    \input{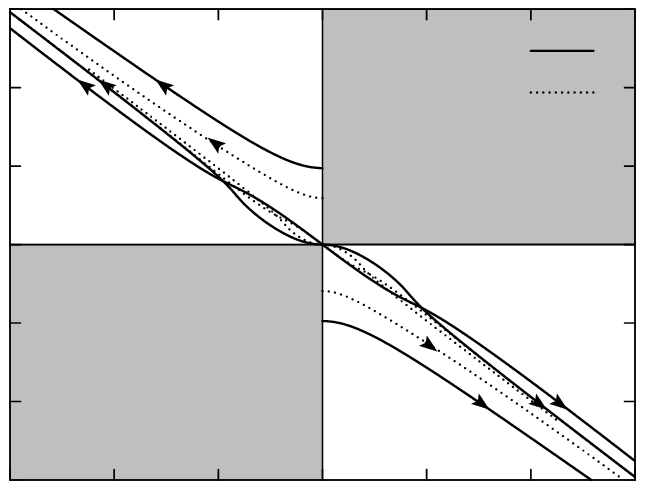}
    \caption{(a) temporal stability analysis; and, (b) spatial stability
    analysis of modified Navier-Stokes equations (grey shaded
            area indicates region of instability).}
    \label{fig:stabilityOettinger}
\end{figure}

We examine temporal stability by solving
(\ref{eq:BNScharacteristics}) numerically for $\omega$ for values of
$k$ in the range $0 \le k < \infty$. Trajectories of $\omega$ are
plotted in the complex plane in
figure~\ref{fig:stabilityOettinger}(a). Stability was tested across
a broad range of $0.2 \leq \Sc \leq 1.0$ and sets of trajectories
are plotted at the two extremes.  In both cases the trajectories all
lie within the region $\omega_{i} \ge 0$, indicating stability for
all $k$.

We then turn to examine spatial stability by solving
(\ref{eq:BNScharacteristics}) numerically for $k$ for values of
$\omega$ in the range $0 \le \omega < \infty$.  Trajectories of $k$
are plotted in the complex plane in
figure~\ref{fig:stabilityOettinger}(b) for the same values of $\Sc$
as before. In both cases, the trajectories do not violate the
stability condition. The results therefore show stable solutions for
the modified Navier-Stokes equations presented in this paper.

\section{The shock structure problem and solution procedure}
\label{sec:shockSimulations}
The shock structure problem presented in this paper concerns the
spatial variation in fluid flow properties across a stationary,
planar, one-dimensional shock in argon.  We define the flow as
moving at an advective speed $u$ in the positive $x$-direction, with
the shock located at $x = 0$; the upstream conditions at $x =
-\infty$ are super/hypersonic and denoted by a subscript `1',
downstream conditions at $x = +\infty$ are denoted by a subscript
`2'.  Shocks were simulated for a range of upstream Mach number %
$1.2 \le \Ma_{1} \le 11.0$ with the problem initialisation and viscosity model
detailed previously\cite{Greenshields&Reese:2007}, but outlined briefly below.

The simulations adopted the same thermodynamic models and
coefficients for argon used in earlier sections.  The diffusive
transport models were those previously described with the ratios of
coefficients specified for argon, notably $\Pr = 2/3$ and $\Sc =
0.76$. The viscosity-temperature relation for argon was
modelled\cite{Greenshields&Reese:2007} by a power law of the form
$\mu = A T^{s}$ using an exponent $s = 0.72$ from independent
experimental data. Since the results for this problem are
historically presented in normalised form, the test problem was
specified conveniently in a nondimensionalised form. The viscosity
coefficient was set to $A = 1$, and in all simulations $p_{1} =
T_{1} = 1$ was specified at the upstream boundary. A gas constant $R
= \gamma^{-1} = 3/5$ was chosen so that $c_{1} = 1$ and, simply,
$u_{1} = \Ma_{1}$ for the particular simulation. At the downstream
boundary, the normal gradient was specified as zero for all
dependent variables except $u_{2}$ whose value was specified using
the Rankine-Hugoniot velocity relation to maintain the shock
stationary and fixed within the domain.

Simulations were performed using our solver, described in detail
elsewhere\cite{Greenshields&Reese:2007}, developed using the open
source Field Operation and Manipulation (OpenFOAM)
software.\cite{openfoam}  A solution domain of 33$\lambda_{M1}$ was
used in all simulations, wide enough to contain the entire shock
structure comfortably.  Initial results were obtained using the
conventional Navier-Stokes equations that converged on a mesh of 800
cells to within 1\% of the solution extrapolated to an infinitely
small mesh size.  The results presented in this paper were produced
with a mesh of 2000 cells, corresponding to a mesh size of $\sim
0.017\lambda_{M1}$. Numerical solutions were executed until they
converged to steady-state, at which point the residuals of all
equations had fallen 5 orders of magnitude from their initial level.

Physical properties such as $\rho$ and $T$ vary continuously through
the shocks from their upstream to their downstream levels over a
characteristic distance of a few mean free paths. Results presented
in this paper are normalised between 0 and 1 and denoted in the
following by the superscript `$^{\star}$', against distance through
the shock, nondimensionalised by $\lambda_{M1}$. Where possible
results are compared with actual experiments
\cite{Steinhilper:1972,Alsmeyer:1976,Torecki&Walenta:1993} rather
than numerical Direct Simulation Monte Carlo (DSMC) data, since the
latter requires certain assumptions relating to the form of the
intermolecular force law.

\section{Results}
\label{sec:results}

\begin{figure}[h!]
    \centering
    \input{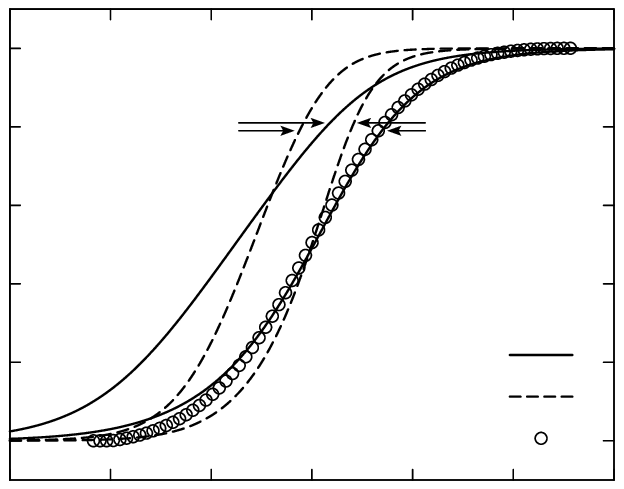}
    \input{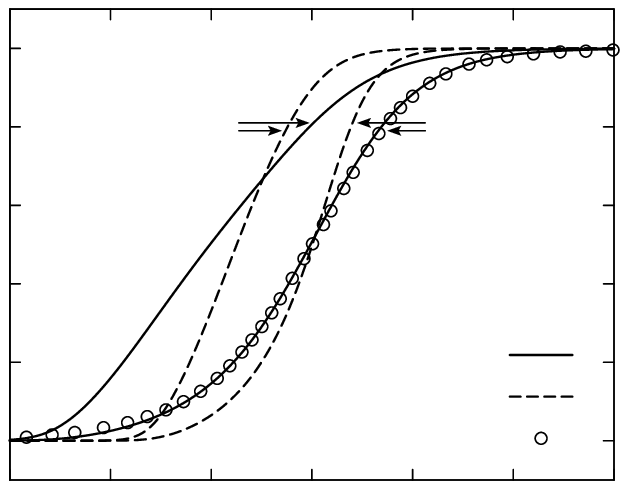}
    \caption{Simulated and experimental profiles of a stationary shock at:
    (a) Mach 2.84; (b) Mach 9.0}
    \label{fig:shockProfiles}
\end{figure}

Figure~\ref{fig:shockProfiles}(a) shows the variation of
$\rho^{\star}$ and $T^{\star}$ through a shock of Mach 2.84
calculated using the Navier-Stokes and modified Navier-Stokes
equations. The experimental density profile of Torecki and
Walenta\cite{Torecki&Walenta:1993} is also shown. It is clear that
the shock layer predicted by the conventional Navier-Stokes
equations is too thin, whereas the modified Navier-Stokes equations
produce good agreement with the experimental data. The main region
of disparity is upstream of the shock layer (left hand side in the
figure) 
where the prediction trails out and is flatter than the experimental
data. Similarly, figure~\ref{fig:shockProfiles}(b) shows the
predicted profiles for a Mach 9.0 shock compared with experimental
density data of Alsmeyer\cite{Alsmeyer:1976}. Again, standard
Navier-Stokes equations produce a shock profile which is too thin
when compared with experiment. However, the modified Navier-Stokes
equations produce excellent agreement with the experimental data.

\subsection{Inverse density thickness}
\label{sec:inverseDensityThickenss}
Apart from direct comparison of calculated and experimental shock
profiles, there are other shock parameters for which experimental
and/or independent numerical data is available.  The principal
parameter is the non-dimensional shock inverse density thickness,
defined as:
\begin{equation}
    L_{\rho}^{-1}
  = \frac{\lambda_{M1}}{\rho_{2} - \rho_{1}}
    |\grad\rho|_{\max} \, .
    \label{eq:inverseDensityThickness}
\end{equation}
Comparing (\ref{eq:Knudsen}) and (\ref{eq:inverseDensityThickness})
it can be seen that, in the absence of a characteristic length scale
$L$ in an unconfined flow, the definition of $\Kn$ requires a
characteristic dimension of a flow structure, in this case the
actual thickness of the shock layer itself. Therefore
$L_{\rho}^{-1}$ has the interesting feature that it represents $\Kn$
for the shock structure case.
\begin{figure}[t]
    \centering
    \input{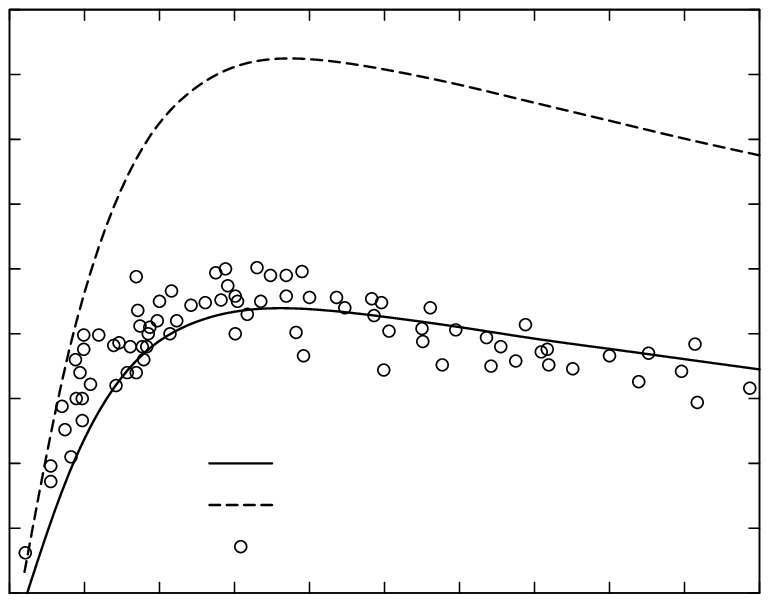}
    \caption{Simulated and experimental inverse density thickness
             ($L_{\rho}^{-1}$) data, versus shock Mach number.}
    \label{fig:MachVsInvShockThickness}
\end{figure}
Alsmeyer\cite{Alsmeyer:1976} reported the most comprehensive
collection of experimental shock data, consisting of his own results
and work published previously.
Figure~\ref{fig:MachVsInvShockThickness} shows $L_{\rho}^{-1}$ for
argon shocks up to Mach 11, comparing simulation results and
experimental data from Alsmeyer\cite{Alsmeyer:1976} and other
sources\cite{Steinhilper:1972,Torecki&Walenta:1993}. The
Navier-Stokes equations predict shocks of approximately half the
measured thicknesses over the entire Mach number range.  As
$L_{\rho}^{-1}$ indicates $\Kn$, this poor agreement is expected
because, over most of this Mach number range, $\Kn \sim 0.2$--$0.3$,
beyond the accepted $\Kn$ limit of application of the Navier-Stokes
equations.
However, results from the modified Navier-Stokes equations match
well with experiment, suggesting that correct prediction of density
gradient has been attained by correct modelling of mass diffusion.

\subsection{Density asymmetry quotient}
\label{sec:densityAsymmetryQuotient}
\begin{figure}
    \begin{minipage}[c]{80mm}
    \centering
    \input{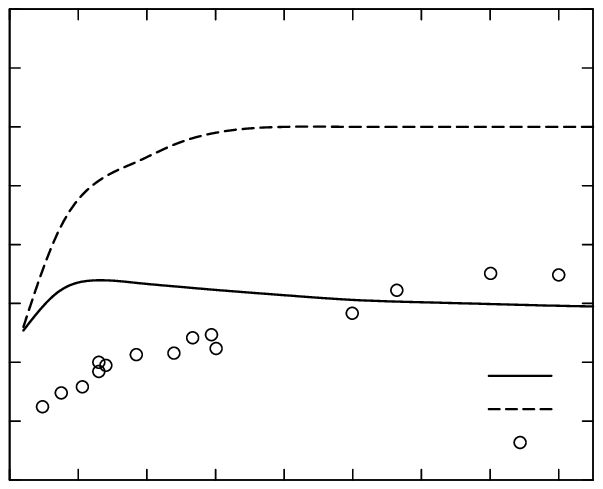}
    \caption{Asymmetry quotient ($Q_{\rho}$) versus shock Mach number.}
    \label{fig:MachVsAsymmetryQuotient}
    \end{minipage}
    \hspace*{3mm}
    \begin{minipage}[c]{80mm}
    \centering
    \input{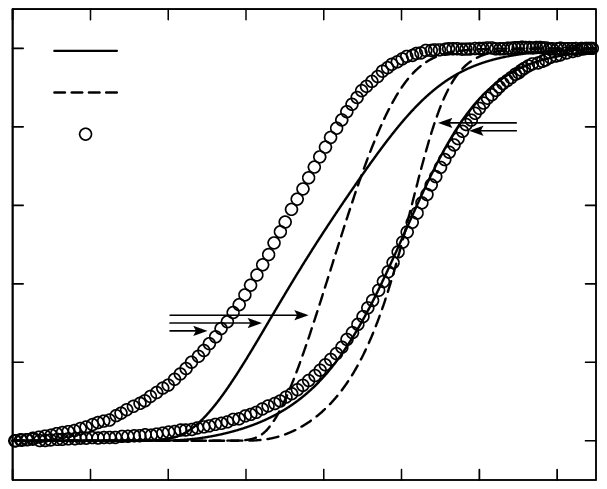}
    \caption{Simulated and DSMC profiles of a stationary shock at Mach 11}
    \label{fig:Mach11profile}
    \end{minipage}
\end{figure}

Agreement of predicted and experimental shock inverse density
thicknesses is not the only measure of the success of a model. As
$L_{\rho}^{-1}$ depends on the density gradient at the profile
midpoint alone, it does not express anything about the overall shape
of the profile. Instead, a second parameter that can be used to
describe the shock profile, and for which experimental data is
available, is the density asymmetry quotient $Q_{\rho}$.  This is a
measure of how skewed the shock density profile is relative to its
midpoint. It is defined for a 1-dimensional profile of normalised
density, $\rho^{\star}$, centred at $\rho^{\star} = 0.5$ on $x = 0$,
as
\begin{equation}
    Q_{\rho}
  =
    \frac%
    {\int_{-\infty}^{0} \rho^{\star}(x)\,\text{d}x}
    {\int_{0}^{\infty} [1 - \rho^{\star}(x)]\,\text{d}x} .
\label{eq:asymmetryQuotient}
\end{equation}
A symmetric shock would consequently have $Q_{\rho} = 1$, but real
shock waves are not completely symmetrical about their midpoint.
First, their general form is skewed a little towards the downstream.
Then, the flattened, diffusive region, that extends upstream of the
shock profile, tends to increase $Q_{\rho}$ with increasing Mach
number. Figure~\ref{fig:MachVsAsymmetryQuotient} shows experimental
data\cite{Alsmeyer:1976} in which $Q_{\rho}$ increases fairly
linearly from $\sim 0.9$ at around Mach 1.5, through unity at around
Mach 2.3, to $\sim 1.15$ at Mach 9.

Results from the Navier-Stokes equations do not agree well with
experimental data: $Q_{\rho} > 1.0$ at Mach 1.2, and rapidly
increases with Mach number with the profile sharpening downstream of
the shock until by Mach 4 it levels off to $Q_{\rho} \approx 1.4$,
compared to $\sim 1.03$ from experiment.  The modified Navier-Stokes
equations similarly overpredict $Q_{\rho}$ at Mach 1.2 but, with
increasing Mach number, $Q_{\rho}$ quickly levels off at $\sim 1.1$
so that, by Mach 6, $Q_{\rho}$ matches well with experiment and the
density profiles are very well predicted, e.g. for the profile at
Mach 9 in figure~\ref{fig:shockProfiles}(b).

\subsection{Temperature-density separation}
\label{sec:temperatureDensitySeparation}
In a shock, the density rises from its upstream value to its
downstream value behind the temperature, due to the finite
relaxation times for momentum and energy transport.  Experimental
data for this phenomenon is scarce due to the difficulty in
measuring temperature profiles, but results from DSMC simulations
provide such data.  Figure~\ref{fig:Mach11profile} shows a
comparison of profiles for a Mach 11 shock from our simulations and
those calculated using DSMC\cite{Lumpkin&Chapman:1991}.  DSMC
clearly predicts a much larger separation distance between density
and temperature profiles than conventional Navier-Stokes equations.
The modifications to Navier-Stokes do increase the
temperature-density separation, though not to the extent of the DSMC
predictions.  The temperature profile predicted by the modified
Navier-Stokes equations is generally less diffusive than that
predicted by DSMC, particularly at the upstream end.

\section{Conclusions}
\label{sec:conclusions}
It is accepted that the conventional Navier-Stokes equations fail to
predict correct shock structures above about Mach 2, where the flow
falls within the intermediate-$\Kn$ regime.  Brenner's modifications
to Navier-Stokes improve the predictions of shock structures
considerably\cite{Greenshields&Reese:2007} but only with the
somewhat arbitrary choice of diffusion coefficient $D_{v} = \nu$
based on an upper limit above which the equations produce unphysical
solutions and, at even higher $D_{v}$, become unstable.

Rather than basing the modifications to Navier-Stokes on the notion
of a diffusive volume flux, we instead present modifications due to
the inclusion of a diffusive mass flux. The resulting set of
governing equations deliver a similar improvement over conventional
Navier-Stokes in reproducing the trends in the experimental and DSMC
data, and in the case of the inverse density thickness produce a
very good match.  The new model uses a \emph{known} coefficient for
self-diffusion of mass for argon and the equation set does not
exhibit the unphysical and unstable behaviour previously observed
with Brenner's modifications.  The new equation set is extremely
easy to solve: it contains none of the higher order derivatives of
the Burnett family of models, nor the second derivative of
density contained within Brenner's original model; indeed, the
addition of a diffusive term to the mass conservation equation makes
the new set arguably easier to solve numerically than the
conventional Navier-Stokes equations themselves.

While it is important not to draw strong conclusions based on just
one test case, the results are generally encouraging.  Our future
aims are: to test this model further on a number of benchmark cases
ranging from high-speed flows encountered in hypersonics to specific
studies of diffusion phenomena such as thermophoresis, and to refine
and develop the models accordingly.

\section*{Acknowledgements}
We would like to thank %
Steve Daley of Dstl Farnborough (UK), %
Henry Weller of OpenCFD Ltd.\ (UK), %
Howard Brenner of MIT (USA) %
and Art Corey of Colorado State University (USA) %
for useful discussions. This work is funded in the UK by the
Engineering and Physical Sciences Research Council under grants
GR/T05028/01 and EP/D007488/1, and through a Philip Leverhulme Prize
for JMR from the Leverhulme Trust.

\bibliographystyle{unsrt}
\bibliography{general,greenshields,hypersonics}

\begin{thebibliography}{10}

\bibitem{Brenner:2004}
H.~Brenner.
\newblock Is the tracer velocity of a fluid continuum equal to its mass
  velocity?
\newblock {\em Physical Review E}, 70:061201, 2004.

\bibitem{Brenner:2005a}
H.~Brenner.
\newblock Kinematics of volume transport.
\newblock {\em Physica A}, 349:11, 2005.

\bibitem{Brenner:2005b}
H.~Brenner.
\newblock {N}avier-{S}tokes revisited.
\newblock {\em Physica A}, 349:60, 2005.

\bibitem{Greenshields&Reese:2007}
C.~J. Greenshields and J.~M. Reese.
\newblock The structure of shock waves as a test of {B}renner's modifications
  to the {N}avier--{S}tokes equations.
\newblock {\em Journal of Fluid Mechanics}, 580:407--429, 2007.

\bibitem{Oettinger:2005}
H.~C. {\"{O}}ttinger.
\newblock {\em Beyond equilibrium thermodynamics}.
\newblock John Wiley and Sons, Hoboken, USA, 2005.

\bibitem{Brenner:2006}
H.~Brenner.
\newblock Fluid mechanics revisited.
\newblock {\em Physica A}, 370:190--224, 2006.

\bibitem{Bird&al:2002}
R.~B. Bird, W.~E. Stewart, and E.~N. Lightfoot.
\newblock {\em Transport phenomena}.
\newblock John Wiley and Sons, Inc., New York, 2002.

\bibitem{Corey&Auvermann:2003}
A.~T. Corey and B.~W. Auvermann.
\newblock Transport by advection and diffusion revisited.
\newblock {\em Vadose Zone Journal}, 2:655--663, 2003.

\bibitem{Mills:1998}
A.~F. Mills.
\newblock The use of the diffusion velocity in conservation equations for
  multicomponent gas mixtures.
\newblock {\em International Journal of Heat and Mass Transfer}, 41:1955--1968,
  1998.

\bibitem{Winn:1950}
E.~B. Winn.
\newblock The temperature dependence of the self-diffusion coefficients of
  argon, neon, nitrogen, oxygen, carbon dioxide, and methane.
\newblock {\em Physical Review}, 80:1024--1027, 1950.

\bibitem{Chapman&Cowling:1970}
S.~Chapman and T.~G. Cowling.
\newblock {\em The mathematical theory of non-uniform gases}.
\newblock Cambridge University Press, Cambridge, UK, third edition, 1970.

\bibitem{Bird:1970}
G.~A. Bird.
\newblock Aspects of the structure of strong shock waves.
\newblock {\em Physics of Fluids}, 13:1172, 1970.

\bibitem{Lofthouse&al:2007}
A.~J. Lofthouse, L.~C. Scalabrin, and I.~D. Boyd.
\newblock Velocity slip and temperature jump in hypersonic aerothermodynamics.
\newblock AIAA Paper 2007-208, 2007.

\bibitem{Zhong&al:1991}
X.~Zhong, R.~W. Mac{C}ormack, and D.~R. Chapman.
\newblock Stabilisation of the {B}urnett equations and application to
  high-altitude hypersonic flows.
\newblock AIAA Paper 91-0770, 1991.

\bibitem{Struchtrup&Torrilhon:2003}
H.~Struchtrup and M.~Torrilhon.
\newblock Regularization of {G}rad's 13 moment equations: {D}erivation and
  linear analysis.
\newblock {\em Physics of Fluids}, 15:2668, 2003.

\bibitem{openfoam}
{O}pen{CFD} {L}td.
\newblock \url{http://www.openfoam.org}, 2004.

\bibitem{Steinhilper:1972}
E.~A. Steinhilper.
\newblock {\em Electron beam measurements of the shock wave structure: Part 1,
  {T}he inference of intermolecular potentials from shock structure
  experiments}.
\newblock PhD thesis, California Institute of Technology, USA, 1972.

\bibitem{Alsmeyer:1976}
H.~Alsmeyer.
\newblock Density profiles in argon and nitrogen shock waves measured by the
  absorption of an electron beam.
\newblock {\em Journal of Fluid Mechanics}, 74:497, 1976.

\bibitem{Torecki&Walenta:1993}
P.~Torecki and Z.~Walenta.
\newblock Private communication.
\newblock {P}olish Academy of Sciences, Warsaw, Poland, 1993.

\bibitem{Lumpkin&Chapman:1991}
F.~E. Lumpkin and D.~R. Chapman.
\newblock Accuracy of the {B}urnett equations for hypersonic real gas flows.
\newblock AIAA Paper 91-0771, 1991.

\end{thebibliography}

\end{document}